\documentclass[letterpaper,prl,twocolumn,nofootinbib,tightenlines,floatfix,preprintnumbers]{revtex4}
\usepackage{graphicx}
\usepackage{amssymb}
\usepackage{bm}
\usepackage{comment}
\usepackage{color}

\newcommand{\be}{\begin{equation}}
\newcommand{\ee}{\end{equation}}
\newcommand{\bea}{\begin{eqnarray}}
\newcommand{\eea}{\end{eqnarray}}

\def\Dslash{D\!\!\!\!/}

\begin{document}

\title{QCD inequalities for hadron interactions}
 
\author{William~Detmold} 
\affiliation{Center for Theoretical Physics,
  Massachusetts Institute of Technology, 
  Cambridge, MA 02139, USA}

\date{\today}

\begin{abstract}
We derive generalisations of the Weingarten--Witten QCD mass inequalities for
particular multi-hadron systems. For systems of any number of identical pseudo-scalar mesons of maximal isospin, these inequalities prove that interactions between the constituent mesons must be repulsive and that no bound states can form in these channels. Similar constraints in less symmetric systems are also extracted. These results are compatible with experimental results (where known) and recent lattice QCD calculations, and also lead to a more stringent bound on the nucleon mass than previously derived, $m_N\ge\frac{3}{2}m_\pi$. 
 \end{abstract}

\preprint{MIT-CTP {4574}}

\maketitle

Analytic relationships between low-energy hadronic quantities are difficult to obtain 
in Quantum Chromodynamics (QCD) because it is a strongly interacting field theory, and only a few such relationships are known.
Consequently, the various  inequalities between hadron masses that have been 
derived by Weingarten \cite{Weingarten:1983uj},  Witten \cite{Witten:1983ut}, and 
(under some assumptions) by Nussinov \cite{Nussinov:1983vh}
have an important place in our understanding of QCD.
The rigorous relations  can be summarised 
by stating that the pion is the lightest colourless state of non-zero isospin \cite{Weingarten:1983uj} 
($m_X\ge m_\pi$ for $X$ being any $I=1$ isospin-charged meson), that the pion electromagnetic mass 
difference $m_{\pi^+} - m_{\pi^0}$ is positive \cite{Witten:1983ut} and that baryons are heavier than 
pions, $m_B\geq m_\pi$ \cite{Weingarten:1983uj,Vafa:1983tf}.   The 
 status of QCD inequalities is reviewed in Ref.~\cite{Nussinov:1999sx}.
The known results concern a relatively small number of  static quantities, and it is important to 
consider whether further relations exist. In this direction, Nussinov and Sathiapalan \cite{Nussinov:1984kr} 
found that in QCD motivated models  there are relationships between scattering lengths in 
various two-particle channels, and  Gupta {\it et al.} \cite{Gupta:1993rn} showed that an 
unphysical combination of $\pi\pi$ interactions is attractive (a physical meaning can be given to
this result in a theory with $N_{f}=4$ degenerate flavours \cite{Sharpe:1992pp}).
  In this letter, we demonstrate   that 
   there are additional rigorous QCD  inequalities
that pertain to the spectrum of physical, multi-hadron systems  and thereby to 
the nature of the corresponding hadronic interactions. As simple examples, we prove that 
there are no bound states in the $I=2$ $\pi^+\pi^+$  or $I=3/2$ $\pi^+K^+$ channels.
As a consequence, we also improve on a previous baryon-meson mass inequality, showing 
that $m_N\ge\frac{3}{2}m_\pi$.

A central observation of Vafa and Witten \cite{Vafa:1983tf} is that the measure of the QCD functional 
integrals that define correlation functions is positive definite in the absence of a $\theta$-term 
or baryon chemical potential (we will ignore these cases throughout). After integrating over the quark degrees of freedom, the measure
can be expressed as
\begin{equation}
d\mu = \prod_{x,\mu,a}d A^a_\mu(x)e^{-S_{YM}[A]}\prod_{f}  \det\left[\Dslash +m_{f}\right] \,,
\end{equation}
where $A_\mu$ represents the gauge field, $\Dslash=\Dslash \;\![A]$ is the fermion Dirac operator, $m_{f}$ is the quark mass of flavour $f$,  and $S_{\rm YM}=\frac{1}{2} \int d^4x{\rm Tr}[F^{\mu\nu}F_{\mu\nu}]$ is 
the Yang-Mills action, and $F^{\mu\nu}=[D^\mu,D^\nu]$. Throughout our discussion, we use a Euclidean metric. Correlation functions involving field operators at $n$ 
points are defined as 
\begin{equation}
\langle {\cal O}(x_1,\ldots,x_n)\rangle =\frac{1}{\cal Z}\int d\mu\;  \hat{\cal O}(x_1,\ldots,x_n)\,,
\end{equation}
where ${\cal Z}=\int d\mu$, and the operator $\hat{\cal O}$ results from the operator ${\cal O}$  
after integration over quark fields. These functional integrals are only defined after the imposition of a regulator, and we assume the use of a regulator that does not spoil positivity \cite{Vafa:1983tf,Weingarten:1983uj}.
As a consequence of the positivity of the measure, field independent relations that are shown to hold for any particular gauge field configuration also hold for the integrated quantity, the corresponding correlation function.
Vafa and Witten used measure positivity to derive the celebrated result  that vector symmetries do not break spontaneously.

In related work, Weingarten \cite{Weingarten:1983uj} considered 
  correlation functions from which meson and baryon masses can be determined, and made use of the 
Cauchy-Schwarz and H\"older inequalities to show that
relationships exist between the corresponding functional integrals. The inequalities show that $m_\pi\le m_X$, and 
$m_{N}\ge \frac{N_f-2}{N_f-3}m_{\pi}$ for a theory with $N_f\ge 6$ flavours. 
Using a further constraint on the spectrum of the inverse
of the Dirac operator, shown to hold in Ref.~\cite{Vafa:1983tf}, 
this latter  constraint was extended to $m_{N}\ge m_{\pi}$, independent of the number of flavours.

Our analysis shares similarities with the above approaches, but also makes use of an eigenvalue 
decomposition of correlation functions. 
We  begin by considering an $I=I_z=n$ many-$\pi^{+}$ correlator of the form
\begin{eqnarray}
\left\langle \Omega \left| \prod_i^{n} u\gamma_5 \bar d(x_i) \prod_j^{n} d\gamma_5 \bar u(y_j) \right|\Omega\right\rangle\,,
\end{eqnarray}
where $|\Omega\rangle$ is the vacuum state and the clusters of points $\{x_{i}\}$ and $\{y_{j}\}$ are taken to be well separated in Euclidean space.
We specify to vanishing total momentum by summing over the spatial components of the $y_{i}$ coordinates
and for simplicity set the temporal components $x_{i}^{4}=0$ $\forall i$ and $y_{j}^{4}=t$ $\forall j$ and allow for 
some of the source locations to be the same (nonzero correlators result provided $4N_c$ or less quark fields are placed at the same point). 
This leads to
\begin{eqnarray}
{\cal C}_{n}&\equiv &C_n({\bf x}_1,\ldots {\bf x}_n;t;{\bf P}=0) \nonumber\\
&=&\left\langle \Omega \left| \prod_i^{n} u\gamma_5 \bar d({\bf x}_i,0) \left[\sum_{{\bf y}}d\gamma_5 \bar u({\bf y},t)\right]^{n} \right|\Omega\right\rangle\,.
\label{corr}
\end{eqnarray}
As shown in Refs.~\cite{Detmold:2010au,Detmold:2012wc}, these correlation functions can be written in terms of 
products of traces of powers of the matrix
\begin{eqnarray}
\mathbf{\Pi}_A=\left(\begin{array}{cccc}
P_{1,1} & P_{1,2} & \cdots & P_{1,N_s} \\
P_{2,1} & \ddots & \ddots & P_{2,N_s}\\
\vdots & \ddots & \ddots & \vdots \\
P_{N_s,1} & \cdots & \cdots & P_{N_s,N_s}
\end{array}\right)\,,
\label{PiDef}
\end{eqnarray}
where  $N_{s}$ is the number of source locations being considered,  the $4N_c\times 4N_c$ blocks are given by
\begin{equation}
P_{i,j}(t)= \sum_{{\bf y}}  
S_u({\bf x}_i,0;{\bf y},t) \gamma_5  S_d({\bf y},t;,{\bf x}_j,0)\gamma_5\,,
\end{equation}
and $S_u$ and $S_d$ are propagators for the  up and down quarks, respectively. The subscript $A$ indicates that
the matrix depends on the background gauge field and $\mathbf{\Pi}_A$ is a matrix of dimension 
$N=4N_cN_{s}$ 
-- in the continuum limit, this can be taken to infinity.

This can be further simplified in the isospin limit where the up and down quark propagators are the same, $S_u=S_d$, and by using the $\gamma_5$
hermiticity of the Dirac operator that implies that 
$\gamma_5 S_d(y,x)\gamma_5 = S_d^\dagger(x,y)$ so
\begin{equation}
P_{i,j}(t)= \sum_{{\bf y}}  
S_u({\bf x}_i,0;{\bf y},t) S_u^\dagger({\bf x}_j,0;{\bf y},t)\,,
\end{equation}
and we see that $\mathbf{\Pi}_A$ is a non-negative definite Hermitian matrix, as are all its diagonal sub-blocks.
In Ref.~\cite{Detmold:2010au}, it was shown that the correlation functions ${\cal C}_j$ for $j\le N$ arise as coefficients 
of the characteristic polynomial 
\begin{eqnarray}
{\cal P}(\alpha)&=&\det(1+\alpha\ \mathbf{\Pi}_A) = \sum_{j=0}^{N} c_j \alpha^j
\end{eqnarray} 
of the matrix $\mathbf{\Pi}_A$.\footnote{There are normalisation differences between the $c_{j}$ and ${\cal C}_{j}$, and for multiple source locations, the $c_j$ are linear combinations of the 
${\cal C}_{j}$ with different numbers of interpolators at each source. The spectrum is common to each term in this linear combination.} Since the roots
of the characteristic polynomial are given by the eigenvalues $\pi_i$ of  $\mathbf{\Pi}_A$, it follows that
\begin{eqnarray}
c_n = \sum_{i_1\neq i_2 \neq \ldots\neq i_n=1}^{N} \pi_{i_1}\pi_{i_2}\ldots \pi_{i_n}\,.
\label{evs}
\end{eqnarray} 
Thus $c_1=\sum_{i=1}^N\pi_i={\rm tr}[\mathbf{\Pi}_A]$, $c_2=\sum_{i=1}^N \sum_{j\ne i =1}^N \pi_i \pi_j$, \ldots, 
$c_N=\pi_1\ldots\pi_N= \det[\mathbf{\Pi}_A]$.
Since these eigenvalues are non-negative, we can bound these correlators by products of the single pion correlator
by relaxing the restrictions on the summation above. That is, 
\begin{eqnarray}
c_n&\leq & 
\sum_{i_1, i_2, \ldots, i_n=1}^{N} \pi_{i_1}\pi_{i_2}\ldots \pi_{i_n}
=\left[\sum_{i=1}^N\pi_i\right]^n = \left[c_1\right]^n\,. \hspace*{6mm}\label{result}
\end{eqnarray}
From this eigenvalue relation, valid on a fixed background gauge configuration, we can 
construct the field independent bound, $c_n - c_1^n \le0$, that holds for all $A_\mu^a$. 
Measure positivity then implies that this relation holds at the level of QCD correlators.
Since the large separation behaviour of  $\langle c_n\rangle $ is governed by the energy of the lowest 
energy eigenstates of the system, $\langle c_n\rangle \sim \exp(-E_n^{(0)} t)$, this implies that 
$E_n^{(0)} \geq n\; E_1^{(0)} = n\, m_\pi$. That is, there are no bound states possible in 
these maximal isospin channels, and further, the two-body interactions in these systems 
are repulsive or vanishing at threshold. This second result follows from the fact that the 
relations derived above are valid in a finite volume where the energy eigenvalues are 
shifted by interactions \cite{Luscher:1986pf,Luscher:1990ux}. Since the scattering 
length is proportional to the negative of the non-negative definite energy shift, it must be 
repulsive or vanishing.\footnote{ We note that  the results hold for common lattice QCD discretisations
such as domain-wall \cite{Kaplan:1992bt} and overlap fermions \cite{Narayanan:1992wx,Narayanan:1994gw} or 
Wilson fermions \cite{Wilson:1974sk} with even numbers of flavours.}

The two-pion results are in accordance with expectations from chiral perturbation theory  ($\chi$PT)
\cite{Weinberg:1966kf,Gasser:1983kx} which predicts at next-to-leading order (NLO)  that 
\begin{eqnarray}
\label{chptpipi}
m_\pi a_{\pi\pi}^{(I=2)} &=& -2\pi \chi \left[1+\chi\left(3 \log\chi -L_{\pi\pi}^{(I=2)}\right)\right]\,,
\end{eqnarray}
 where $\chi=[m_{\pi}/4\pi f_{\pi}]^{2}$,  $f_\pi$ is the pion decay constant, and $L_{\pi\pi}^{(I=2)}$ is a particular combination of low energy constants (LECs) renormalised at scale $\mu=4\pi f_{\pi}$.
At tree level, this expression is universally negative, and at NLO it remains negative given the phenomenological constraints on $L_{\pi\pi}^{(I=2)}$.
However, the  bounds derived above are statements directly about QCD and do not rely on a 
chiral expansion, and  in fact provide a fundamental constraint on $L_{\pi\pi}^{(I=2)}$ (the use of single particle QCD inequalities to constrain $\chi$PT is discussed in Refs.~\cite{Bar:2010ix,Hansen:2011kk}). The $\pi\pi$ scattering phase shifts can be experimentally extracted from studies of kaon decays \cite{Pislak:2001bf,Pislak:2003sv,Batley:2005ax} and the lifetime of pionium \cite{Adeva:2005pg}, but the direct constraints of the $I=2$ channel are relatively weak. A chiral and dispersive analysis of experimental data nevertheless allows for a precise extraction 
\cite{Colangelo:2001df}, giving $m_\pi a_{\pi\pi}^{(I=2)}=-0.0444(10)$
 and lattice QCD calculations \cite{Yamazaki:2004qb,Beane:2005rj,Beane:2007xs,Feng:2009ij,Yagi:2011jn,Beane:2011sc,Fu:2013ffa} are in agreement. The sign implies that these results are  concordant with the QCD inequalities derived here.

As a corollary, having shown that the $(\pi^+)^n$ systems do not bind, we can follow the discussion of  Ref.~\cite{Nussinov:1999sx} and strengthen the nucleon mass bound of Weingarten  to $m_N\ge \frac{3}{2}m_\pi$. This improves on the bounds of Refs.~\cite{Weingarten:1983uj,Nussinov:1984kr,Cohen:2003ut} as it applies for arbitrary $N_{f}$ and $N_{c}$ and the inequality directly involves the pion mass.
Furthermore, less complete modifications of the restricted sums in Eq.~(\ref{evs}) show also that $E_n^{(0)} \ge E_{n-j}^{(0)}+ E_j^{(0)}$ for all $j> n$. This then implies that the $I=3$, $\pi^+\pi^+\pi^+$ interaction is separately repulsive at threshold, as are the $I=n$, $(\pi^+)^n$ interactions. In principle, the form of these interactions could be computed  in the chiral expansion, and the 
constraints derived here would bound  the LECs that enter. Lattice calculations
show that the $\pi^+\pi^+\pi^+$ interaction is indeed repulsive  \cite{Beane:2007es,Detmold:2008fn}.

\begin{figure}
\centering
\includegraphics[width=0.8\columnwidth]{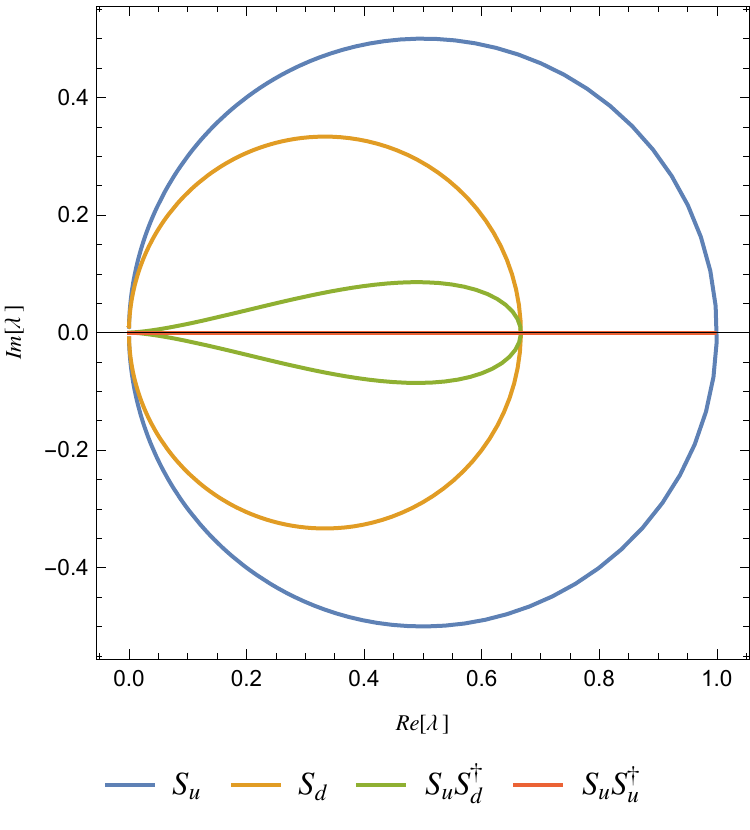}
\caption{Eigenvalues of $S_u$, $S_d$, $S_u S_u^\dagger$ and $S_u S_d^\dagger$ for $m_u=1$ and $m_d=1.5$.}
\label{f1}
\end{figure}
The inequalities above concern identical pseudoscalar mesons formed from quarks of 
equal mass, but can be generalised in a number of ways. In particular, these inequalities 
can be extended to the case of unequal quark masses; thereby analogous 
results can be derived multiple pion systems away from the isospin limit. Further, by defining 
\begin{equation}
K_{i,j}(t)= \sum_{{\bf y}}  
S_u({\bf x}_i,0;{\bf y},t) S_s^\dagger({\bf y},t;,{\bf x}_j,0)\,,
\end{equation}
where $S_s(x,y)$ is the strange quark propagator, 
in addition to $P_{i,j}$, correlators containing both $\pi^+$ and 
$K^+$ mesons can be studied. The matrix $\mathbf{K}_A$ can be constructed from the 
$K_{i,j}$ sub-blocks analogously to Eq.~(\ref{PiDef}). To see how these generalisations arise, we 
need to examine the spectrum of the relevant matrices. 
If we denote the eigenvalues and eigenfunctions of the Dirac operator
as $\lambda_i$ and $v_i$ respectively, that is $\Dslash\ v_i = \lambda_i v_i$, we can  decompose the quark propagators as
\begin{eqnarray}
S_q = \sum_i \frac{v_i v_i^\ast}{\lambda_i + m_q}
 \equiv\sum_i \sigma_i^{(q)} v_i v_i^\ast\,,
\end{eqnarray}
and the matrix $\mathbf{\Pi}_A$  as
\begin{eqnarray}
\mathbf{\Pi}_A&\equiv& \sum_i \pi_i v_i v_i^\ast\nonumber
=\sum_{i,j} \frac{v_i v_i^\ast}{\lambda_i + m_u}\left(\frac{v_j v_j^\ast}{\lambda_j + m_d}\right)^\dagger   \\
&=& \sum_i \frac{v_i v_i^\ast (-\lambda_i^2+m_um_d +\ \lambda_i (m_u-m_d))}{(m_u^2 -\lambda_i^2 )(m_d^2-\lambda_i^2 )}\,,
\hspace*{3mm}
\end{eqnarray}
with a similar expression for $\mathbf{K}_A$ (in the second equality for $\mathbf{\Pi}_A$, 
we have used completeness as we are integrating over the position of the sink in defining $P_{i,j}$).
Because of the spectral properties of the Dirac operator ($\lambda_i\in {\mathbb I}$, and $\{\lambda_i,\lambda^\ast_i\}$ both eigenvalues), the 
eigenvalues of quark propagators, $\sigma_i^{(q)}$, fall on circles 
(centre $(1/(2m_q),0)$, radius $1/(2m_q)$) in the complex plane. 
For the matrix $\mathbf{\Pi}_A$ in the isospin limit,
we immediately see that the eigenvalues are real and non-negative as stated above, occupying the interval $[0,1/m_q]$. 
Away from the isospin limit, $\mathbf{\Pi}_A$ and $\mathbf{K}_A$ have eigenvalues, 
denoted $\pi_i$ and $\kappa_i$ respectively,  that occur 
in complex conjugate pairs with non-negative real parts and imaginary parts that are 
proportional to the mass splitting $|m_1-m_2|$. The locii of these eigenvalues are shown in 
Fig.~\ref{f1} for exemplary masses.

\begin{figure}
\centering
\includegraphics[width=0.8\columnwidth]{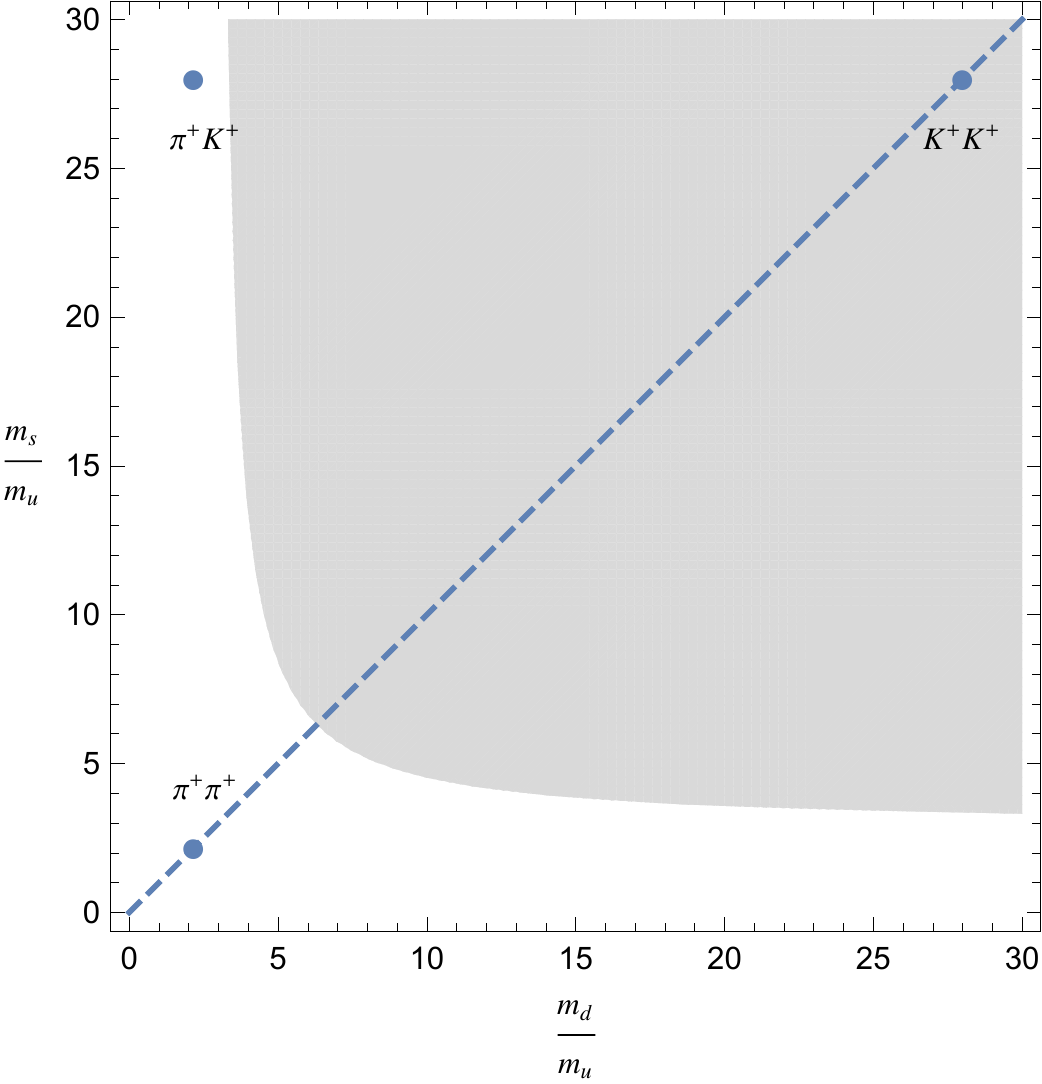}
\caption{Grey shading indicates the region of non-positivity of the integrand in Eq.~(\protect{\ref{mixedint}}). Also shown are relevant physical mass ratios for $\pi^+\pi^+$ at $m_d\ne m_u$ and $\pi^+K^+$ and $K^+K^+$.}
\label{f2}
\end{figure}

The properties discussed above are enough to show 
that even in the less symmetric cases mentioned above, the generalisations of the eigenvalue inequality  used in 
Eq.~(\ref{result}) still hold, at least for certain quark mass ratios in systems containing up to $n=8$ 
particles (for example $\pi^+\pi^+\pi^+K^+$) where we have explicitly checked.\footnote{We expect  that these results hold for all $n$ and all mass ratios, but have been unable to prove the necessary relations.}
To see this, we reconsider the eigenvalue sums that occur in the expressions
 for correlators\footnote{The correlators for $j$ $\pi^+$s and $k$ $K^+$s 
can be constructed from the expansion of $\det(1+\alpha(\mathbf{\Pi}_A+\beta\mathbf{K}_A)$ 
as discussed in Ref.~\cite{Detmold:2011kw}.}
with the quantum numbers of $(\pi^+)^j(K^+)^{n-j} $, denoted $c_{j,n-j}$.
As the simplest example we consider
\begin{eqnarray}
c_{1,1}\sim\sum_i\sum_{j\ne i}\pi_i\kappa_j 
&=& \sum_{i,j}\pi_i\kappa_j  - \sum_{i}\pi_i\kappa_i\,,
\end{eqnarray}
and shall show that the last sum is positive. This is most easily approached in the $N\to\infty$ limit in which the eigenvalue sums become continuous integrals. To make our notation simpler, we replace $\lambda\to i \lambda_R$ with $\lambda_R\in\mathbb{R}$ and subsequently drop the subscript. In this case, defining
\begin{eqnarray}
f_{a,b}(\lambda)= \frac{\lambda^2+m_am_b +i\ \lambda (m_a-m_b))}{(\lambda^2 + m_a^2)(\lambda^2 + m_b^2)}\,,
\end{eqnarray}
and $\pi(\lambda)=f_{u,d}(\lambda)$ and $\kappa(\lambda)=f_{u,s}(\lambda)$,
we can replace $\sum_{i}\pi_i\kappa_i$  by
\begin{eqnarray}
\label{mixedint}
 \int_{-\infty}^\infty {\cal D}\lambda \;\pi(\lambda)\;\kappa(\lambda) 
 \,,
\end{eqnarray}
where the measure ${\cal D}\lambda\equiv d\lambda\; \rho(\lambda)$ is weighted by the spectral density of the Dirac operator, $\rho(\lambda)$. Since the spectral density is non-negative, a non-negative integrand results in a non-negative integral. However, the integrand above is only positive 
definite for some ranges of the ratios ${m_d}/{m_u}$ and ${m_s}/{m_u}$ as is shown for this case in Fig.~\ref{f2}. If the mass ratios are in the allowed region, then $c_{1,1}\le c_{1,0}c_{0,1}$ and through the 
same logic that we employed for $I=2$ $\pi\pi$ systems, we see that $E_{\pi^+K^+}\ge m_{\pi^+}+m_{K^+}$, so   $I=3/2$ $\pi^+ K^+$ scattering 
can not result in bound states. {This result is in agreement with lattice  calculations \cite{Beane:2006gj,Detmold:2011kw,Fu:2011wc,Sasaki:2013vxa}.} 
Outside these parameter ranges, the 
integral has negative contributions at intermediate $\lambda$ but is positive at large $\lambda$; given the expectations of the behaviour of the spectral density, $\rho(\lambda)\sim V \lambda^3$ 
for large $\lambda$,  this suggests that the integral is always positive in Eq,~(\ref{mixedint}), but this cannot be proven rigorously. For the important cases  of $\pi^+\pi^+$ at $m_d\ne m_u$ and $\pi^+K^+$, the physical mass ratios \cite{Aoki:2013ldr} are such that the proof is complete, but for example for  $I=1$ $K^+K^+$ or $D^+D^+$, the mass ratios are such that the proof fails.

In a more complicated case, such as $c_{3,1}$, the subtractions are more involved,
\begin{eqnarray}
c_{3,1}&\sim&\sum_i\sum_{j\ne i}\sum_{k\ne i,j}\sum_{l\ne i,j,k}\pi_i\pi_j \pi_k\kappa_l  \\
&\hspace*{-4mm}=&\hspace*{-4mm} \sum_{ i,j,k,l}\pi_i\pi_j \pi_k\kappa_l - 3
\sum_i\sum_{j\ne i}\sum_{k\ne i,j} (\pi_i^2\pi_j\kappa_k +  \pi_i\pi_j\pi_k\kappa_i)
\nonumber \\
&&\hspace*{-4mm}-\sum_i\sum_{j\ne i}(\pi_i^3\kappa_j +3\pi_i^2\pi_j \kappa_i+3\pi_i^2\pi_j \kappa_j)
-\sum_i \pi_i^3\kappa_i \nonumber\\
&\hspace*{-4mm}=&\hspace*{-4mm} \sum_{ i,j,k,l}\pi_i\pi_j \pi_k\kappa_l  - \Big\{3 \sum_{ i,j,k}(\pi_i^2\pi_j \kappa_k +\pi_i\pi_j\pi_k\kappa_i) 
\nonumber \\
&&\hspace*{-4mm}-\sum_{i,j}(2\pi_i^3\kappa_j+6\pi_i^2\pi_j\kappa_i+3\pi_i^2\pi_j\kappa_j)
+6\sum_i \pi_i^3\kappa_i\Big\}\,.
\nonumber
\end{eqnarray}
However, by again taking the continuous limit and writing the eigenvalue sums as (multiple) integrals, the term in the braces can be proven to be positive for certain values of ${m_d}/{m_u}$ and ${m_s}/{m_u}$, thereby showing  $E_{\pi^+\pi^+\pi^+K^+}\ge 3m_{\pi^+}+m_{K^+}$. The region
of guaranteed positivity varies with the number of pions and kaons in the system, but a region exists for all $c_{j,k}$.

As a further generalisation, we may consider modified correlators where we replace some of the $\gamma_5$ matrices in Eq.~(\ref{corr}) by other Dirac structures.
We can then use the Cauchy-Schwartz inequality to derive the related results that the energies of arbitrary $J^P$ states with $I=I_z=n$ are bounded from below by 
$n\, m_\pi$ in the same manner in which Weingarten \cite{Weingarten:1983uj} showed that $m_X \ge m_\pi$. This does not prohibit bound state formation if the quantum numbers prohibit an $n$ $\pi^+$ state in the given channel (for example $\rho^+\rho^+\rho^+$ with $J^P=3^-$), 
but limits the amount of binding that is possible.

In summary, we have shown that the hadron mass inequalities previously derived in QCD have an infinite set of analogues for  multi-hadron systems that constrain the nature of the  interactions between the constituent hadrons. 
These results provide important constraints on phenomenological, and lattice QCD studies of hadron interactions.
The scope of the techniques used to derive the 
original hadron mass inequalities and the new techniques introduced here is more general than the two-point 
correlation functions considered so far, and there are a number of extensions that may
be pursued productively.

\acknowledgments{
This work was supported by the US Department of Energy Early Career Research Award DE-SC0010495
 and the Solomon Buchsbaum Fund at MIT. 
The author is grateful to M.~J.~Savage, S. Sharpe and B. Tiburzi for  discussions and comments.}

\bibliography{WVW}
  
\end{document}